\documentclass[a4paper,11pt]{article}
\pdfoutput=1 

\usepackage{amsmath}                    
\usepackage[bbgreekl]{mathbbol}
\DeclareSymbolFontAlphabet{\mathbbl}{bbold}
\usepackage{amsfonts,amssymb,amsthm,mathtools}    
\DeclareSymbolFontAlphabet{\mathbbm}{bbold}
\DeclareSymbolFontAlphabet{\mathbb}{AMSb}%
\usepackage{tensor,mathrsfs}
\usepackage[utf8]{inputenc} 
\usepackage[T1]{fontenc}
\usepackage{geometry} 				
\usepackage{lmodern}
\usepackage{enumerate}
\usepackage{titlesec}
\usepackage{bm}
\usepackage[dvipsnames]{xcolor}            
\usepackage[customcolors]{hf-tikz}
\usepackage{array}
\usepackage{authblk}
\usepackage[numbers,sort&compress]{natbib}
\usepackage{scalerel}
\usepackage{tensor}
\usepackage{jheppub}
\usepackage{accents}
\usepackage{todonotes}
\usepackage{accents}

\newcommand{\corurl}{BrickRed}  \newcommand{\corcite}{red}
\newcommand{\corlink}{blue}    \newcommand{\corfile}{black}

\hypersetup{
	linktocpage,
	colorlinks,
	urlcolor=\corurl,	
	citecolor=\corcite,
	linkcolor=\corlink,
	filecolor=\corfile,
	pdfnewwindow,
	pdftitle={Palatini gravity},
	pdfauthor={FB JMB VV ESV},
	pdfsubject={CPS}}
\usepackage{todonotes}
\usetikzlibrary{matrix}

\def\ledgee{{\setbox0\hbox{\ensuremath{\mathrel{\cdot}}}\rlap{\hbox to \wd0{\hss\ensuremath\wedge\hss}}\box0}}

\usepackage{graphicx}
\makeatletter
\newcommand\rrule[3][0pt]{%
	\ifdim#2>#3\math@hrule[#1]{#2}{#3}\else\math@vrule[#1]{#2}{#3}\fi}
\newcommand\math@hrule[3][0pt]{%
	\gdef\mystery@factor{0.07}%
	\@tempdima=#3%
	\rule[#1]{0pt}{#3}
	\raisebox{.5\@tempdima+#1}{%
		\makebox[#2][l]{\kern-.5\@tempdima\@@mathrule{#2}{#3}}}%
}
\newcommand\math@vrule[3][0pt]{%
	\gdef\mystery@factor{0.0}%
	\@tempdima=#2%
	\rule[#1]{0pt}{#3}
	\raisebox{-.0\@tempdima+#1}{%
		\kern0.5\@tempdima%
		\rotatebox{90}{\kern-0.5\@tempdima\makebox[#3][l]{\@@mathrule{#3}{#2}}}%
		\kern0.5\@tempdima}%
}
\def\@@mathrule#1#2{%
	\@tempdimb=#2%
	\@tempdima=\dimexpr#1-\mystery@factor\@tempdimb
	\pdfliteral{%
		q []0 d %
		1 J 
		\strip@pt\@tempdimb\space w \strip@pt\@tempdimb\space 0 m %
		\strip@pt\@tempdima\space 0 l S Q }}
\makeatother

\graphicspath{{./font/}{./}}

\newcommand{\dd}{\mathchoice
	{\mathbbm{d}\rrule{.087ex}{1.605ex}\hspace*{0.15ex}} 
	{\mathbbm{d}\rrule{.087ex}{1.605ex}\hspace*{0.15ex}} 
	{\mathbbm{d}\rrule{.08ex}{1.125ex}\hspace*{0.15ex}}  
	{\mathbbm{d}\rrule{.06ex}{.8ex}\hspace*{0.15ex}}     
}

\newlength{\alturaL}

\newlength{\alturaO}
\newcommand{\OOmega}{\includegraphics[height=\alturaO]{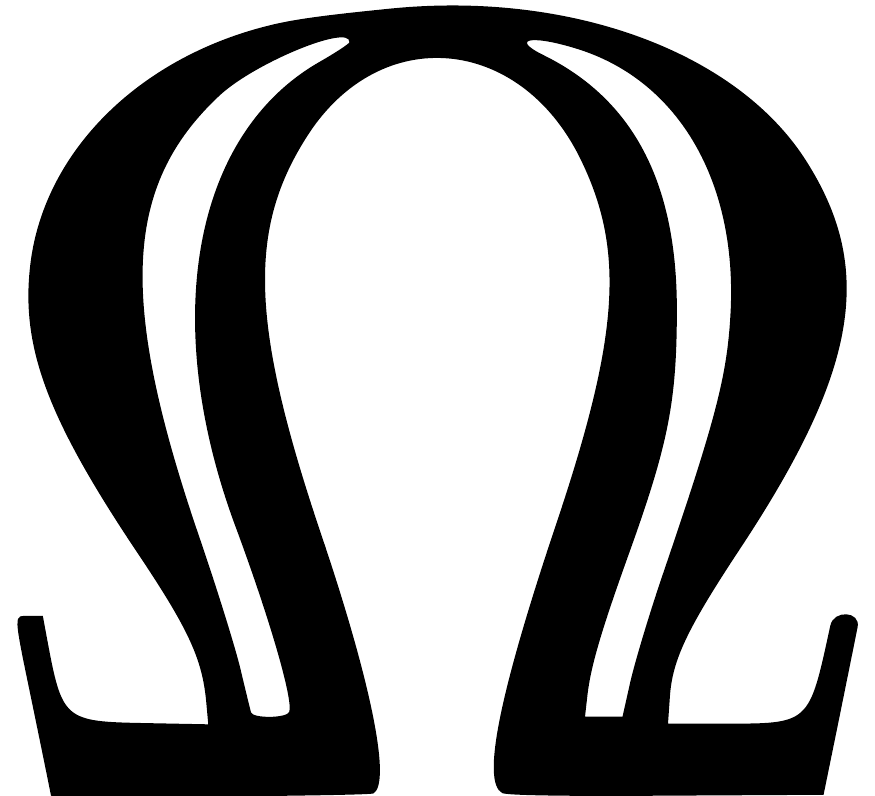}}

\newlength{\alturaI}

\newlength{\alturaJ}

\newlength{\alturaX}

\newcommand{\vol}{\mathrm{vol}}

\renewcommand{\d}{\mathrm{d}}
\newcommand{\LC}[1]{\accentset{\circ}{#1}}

\usepackage{titlesec}
\usepackage{slashed}

\title{Palatini gravity with nonmetricity, torsion, and boundaries in metric and connection variables}

\author[c,d]{J. Fernando Barbero G.}   
\author[a,c]{Juan Margalef-Bentabol}
\author[b,c]{Valle Varo}
\author[b,c]{Eduardo J.S.~Villaseñor}

\emailAdd{fbarbero@iem.cfmac.csic.es}
\emailAdd{juanmargalef@psu.edu}
\emailAdd{valle@cvb.es}
\emailAdd{ejsanche@math.uc3m.es}

\affiliation[a]{Institute for Gravitation and the Cosmos and Physics Department. Penn State
	University. PA 16802, USA.
	\vspace*{1ex} \mbox{}}
\affiliation[b]{Departamento de Matemáticas, Universidad Carlos III de Madrid. Avda. de la
	Universidad 30, 28911 Leganés, Spain.
	\vspace*{1ex} \mbox{}}
\affiliation[c]{Grupo de Teorías de Campos y Física Estadística. Instituto Gregorio Millán (UC3M).
	Unidad Asociada al Instituto de Estructura de la Materia, CSIC, Madrid, Spain.}
\affiliation[d]{Instituto de Estructura de la Materia, CSIC, Serrano 123, 28006, Madrid, Spain}

\abstract{We prove the equivalence in the covariant phase space of the metric and connection formulations for Palatini gravity, with nonmetricity and torsion, on a spacetime manifold with boundary.  To this end, we will rely on the cohomological approach provided by the relative bicomplex framework. Finally, we discuss some of the physical implications derived from this equivalence in the context of singularity identification through curvature invariants.}


\arxivnumber{}

\titlespacing\section{0pt}{0pt plus 4pt minus 2pt}{0pt plus 2pt minus 2pt}
\titlespacing\subsection{0pt}{8pt plus 4pt minus 2pt}{0pt plus 2pt minus 2pt}
\titlespacing\subsubsection{0pt}{8pt plus 4pt minus 2pt}{0pt plus 2pt minus 2pt}

\begin{document}

\maketitle


\section{\label{sec:INTRODUCTION}INTRODUCTION}

The study of general relativity on manifolds with boundary is of great importance as these can be used to account for the asymptotics of the gravitational field and the presence of horizons \cite{Ashtekar:2018lor,Ashtekar:2004cn,Wieland:2020gno}. The original formulation of general relativity relied on a Lorentzian metric as the fundamental field. However, it is also possible to use tetrads for the same purpose. This is important from a physical point of view for several reasons. First, the use of tetrads provides a natural way to incorporate fermion fields into the theory. Second, the tetrad formalism leads naturally to the Ashtekar formulation, which is one of the main avenues for the quantization of gravity \cite{Ashtekar:2004eh,BarberoG.:2020tmm}. Finally, local Lorentz gauge invariance plays a crucial role in the tetrad formalism, which may introduce significant differences in the treatment of the theory compared with the metric formulation.  Within both frameworks, there are several possible choices for the action that can be classified according to their independent basic field variables. Palatini theories are those in which the actions are written either in terms of a metric and a connection (metric formalism) or a frame and a spin connection (tetrad formalism) \cite{Hehl1976,Hehl:1994ue,Olmo:2011uz,Afonso:2017bxr}.\vspace*{2ex}

The present paper attempts to unify and expand certain aspects of previous works \cite{Julia:1998ys,Banerjee2010,Pons2012,Barnich:2016rwk,Corichi:2016zac,Latorre2018, Chakraborty:2018qew,DePaoli:2018erh,Oliveri2020,Oliveri:2020xls,Barnich:2020ciy}, focusing on the comparison of both formulations and the treatment of boundaries. In this regard, we would like to highlight the pioneering work by Obukhov \cite{Obukohv1987}, where he introduced the appropriate surface terms for Palatini gravity. Here we generalize those results and improve on the variational treatment of the problem by relying on the recently proposed CPS algorithm \cite{CPS}, which provides a clean, consistent, and ambiguity-free procedure to obtain the solution spaces, the presymplectic forms canonically associated with the actions, and some relevant charges.\vspace*{2ex} 

The equivalence of the metric-Palatini and tetrad-Palatini formalisms will be proven in two steps. First, we will obtain a precise description of the solution spaces for both theories which will allow us to map them appropriately. Second, we will show that the presymplectic forms given by the CPS algorithm are equivalent in both cases.\vspace*{2ex}

In the following we consider a 4-dimensional spacetime manifold $M$ diffeomorphic to $\Sigma\times\mathbb{R}$, where $\Sigma$ is a 3-dimensional manifold with boundary $\partial\Sigma$ (possibly empty).  We will refer to $\partial_LM\cong\partial\Sigma\times\mathbb{R}$ as the \textit{lateral boundary} of $M$ and restrict ourselves to the open set of metrics making $\partial_LM$ time-like. A few words on notation: Greek letters will denote abstract indices for tensorial objects in $M$ and barred Greek indices will denote tensors on $\partial_LM$ (quite often the object itself will also carry an overbar). The inclusion map will be denoted as $\jmath:\partial_L M\hookrightarrow M$ and its tangent map as $\jmath^\alpha_{\overline{\alpha}}$.

\section{\label{sec:METRICPALATINI}METRIC PALATINI}
Given a connection $\widetilde{\nabla}$, we define its torsion, Riemann, and Ricci tensors as
\begin{align*}
    &\tensor{\widetilde{\mathrm{T}}\mathrm{or}}{^\alpha_\mu_\nu}(\d\phi)_\alpha =-[\widetilde{\nabla}_{\!\mu},\widetilde{\nabla}_{\!\nu}]\phi\,,\\
    &\tensor{\widetilde{\mathrm{R}}\mathrm{iem}}{^\alpha_\beta_\mu_\nu}Z^\beta =([\widetilde{\nabla}_{\!\mu},\widetilde{\nabla}_{\!\nu}]+\tensor{\widetilde{\mathrm{T}}\mathrm{or}}{^\beta_\mu_\nu}\widetilde{\nabla}_{\!\beta})Z^\alpha\,,\\
    &\widetilde{\mathrm{R}}\mathrm{ic}_{\beta\nu}:=\tensor{\widetilde{\mathrm{R}}\mathrm{iem}}{^\mu_\beta_\mu_\nu}\,. 
\end{align*}
If we endow $M$ with a connection $\widetilde{\nabla}$ and a metric $g$, we have the nonmetricity tensor, the $(g,\!\widetilde{\nabla})$-scalar-curvature,  the  $(g,\!\widetilde{\nabla})$-extrinsic-curvature of $\partial_LM$, and its trace
\begin{align*}
    &\widetilde{M}_{\alpha\beta\gamma}:=\widetilde{\nabla}_{\!\alpha}g_{\beta\gamma}\,,
    &&\widetilde{R}:=g^{\alpha\beta}\widetilde{\mathrm{R}}\mathrm{ic}_{\alpha\beta}\,,\\
    &\widetilde{K}_{\overline{\alpha}\overline{\beta}}:=\frac{1}{2}\jmath^\alpha_{\overline{\alpha}}\jmath^\beta_{\overline{\beta}}\Big(\widetilde{\nabla}_{\!\alpha}\nu_\beta+g_{\alpha\gamma}\widetilde{\nabla}_{\!\beta}\nu^\gamma\Big),
    &&\widetilde{K}:=\overline{g}^{\overline{\alpha}\overline{\beta}}\widetilde{K}_{\overline{\alpha}\overline{\beta}}\,,
\end{align*}
where  $\overline{g}:=\jmath^*g$ is the induced metric, $\nu^\alpha$ the outward unit vector field, and $\nu_\beta=g_{\beta\gamma}\nu^\gamma$. Notice that $\widetilde{R}$ and $\widetilde{K}_{\overline{\alpha}\overline{\beta}}$ are (non-standard) generalizations of the $g$-scalar $\LC{R}$ and $g$-extrinsic curvature $\LC{K}$ defined by the $g$-Levi-Civita connection $\LC{\nabla}$. Moreover, in general, $\widetilde{K}_{\overline{\alpha}\overline{\beta}}$ is not symmetric if the nonmetricity is different from zero.\vspace*{2ex}

Given two connections $\nabla$ and $\widetilde{\nabla}$, their difference is a $(2,1)$-tensor $Q\equiv\widetilde{\nabla}-\nabla$. For a $(1,1)$-tensor $\tensor{S}{^\beta_\gamma}$ we have
\[(\widetilde{\nabla}_{\!\alpha}-\nabla_{\!\alpha})\tensor{S}{^\beta_\gamma}=\tensor{Q}{^\beta_\alpha_\mu}\tensor{S}{^\mu_\gamma}-\tensor{Q}{^\mu_\alpha_\gamma}\tensor{S}{^\beta_\mu}\]
and analogously for higher order objects. Observe that if we choose a fiducial connection, usually the $g$-Levi-Civita one, there is a bijection between connections $\widetilde{\nabla}$ and $(2,1)$-tensors $Q$. Working with tensors is usually easier, as they form a vector space while connections form an affine space. Thus, in the following, we will use the variables $(g,Q)$ instead of the equivalent ones $(g,\!\widetilde{\nabla})$.\vspace*{2ex}

We will now use the CPS algorithm \cite{CPS}, which essentially consists in introducing a pair of bulk and boundary Lagrangians, compute the variations, extract the equations of motion and symplectic potentials, and get the presymplectic form on the space of solutions. The power of this method lies in its cohomological nature which renders it ambiguity-free: we can pick any representative Lagrangians and symplectic potentials to describe the solution spaces and compute the presymplectic form.

\subsection{The action}

We consider actions of the form
\[
\mathbb{S}=\int_ML-\int_{\partial M}\overline{\ell}\,,
\]
for some bulk Lagrangian $L$ and some boundary Lagrangian $\overline{\ell}$. The metric-GR action and its generalization, known as metric-Palatini, are respectively given by the Lagrangian pairs
\begin{align*}
    &L^{(m)}_{\mathrm{EH}}(g):=\big(\LC{R} - 2\Lambda \big)\vol_g\,, & &\overline{\ell}^{(m)}_{\mathrm{GHY}}(g):=-2\LC{K} \, \vol_{\overline{g}}\,,\\
    &L^{(m)}_{\mathrm{PT}}(g, Q):= \big(\widetilde{R} - 2\Lambda \big)\vol_g\,, & &\overline{\ell}^{(m)}_{\mathrm{PT}}(g, Q):=-2\widetilde{K} \, \vol_{\overline{g}}\,.
\end{align*}
It is straightforward to rewrite the Palatini Lagrangians as those of standard GR plus a coupling term
\begin{align*}
    &L^{(m)}_{\mathrm{PT}}(g,Q)=L^{(m)}_{\mathrm{EH}}(g)+L^{(m)}_{\mathrm{CP}}(g,Q)\,,\\
    &\overline{\ell}^{(m)}_{\mathrm{PT}}(g,Q)=\overline{\ell}^{(m)}_{\mathrm{GHY}}(g)+\overline{\ell}^{(m)}_{\mathrm{CP}}(g,Q)\,,
\end{align*}
where
\[
    \begin{array}{l}L^{(m)}_{\mathrm{CP}}(g,Q):=\big(C_{\lambda}A^{\lambda} - \tensor{Q}{^\alpha^\beta_\lambda}\tensor{Q}{^\lambda_\alpha_\beta}\big) \vol_{g} + \d \big(\iota_{\vec{A} - \vec{C}}\vol_{g}\big)\,,\\[1ex]
      \overline{\ell}^{(m)}_{\mathrm{CP}}(g,Q): =\jmath^{*}\big(\iota_{\vec{A}-\vec{C}}\vol_{g} \big)\,,\end{array}\qquad
     \begin{array}{l}A^\alpha:=g^{\beta\gamma}\tensor{Q}{^\alpha_\beta_\gamma}\,,\\ B_\beta:=\tensor{Q}{^\mu_\beta_\mu}\,,\\ C_\gamma:=\tensor{Q}{^\mu_\mu_\gamma}\,.\end{array}
\]
As mentioned before, the results given by the CPS algorithm do not depend on the choice of Lagrangians as long as they define the same action (the pairs are equal up to a \emph{relative} exact form, see \cite{CPS} for more details). In the present case it is easy to see that
\begin{align*}
    & \big(L^{(m)}_{\mathrm{PT}},\overline{\ell}^{(m)}_{\mathrm{PT}}\big)\!=\!\big(L^{(m)}_{\mathrm{EH}},\overline{\ell}^{(m)}_{\mathrm{GHY}}\big)\!+(\hat{L}^{(m)}_{\mathrm{CP}},0)\!+\underline{\d}(\iota_{\vec{A}-\vec{C}}\vol_g,0)\,, \\
  &\hat{L}^{(m)}_{\mathrm{CP}}:=L^{(m)}_{\mathrm{CP}}-\d(\iota_{\vec{A}-\vec{C}}\vol_g)\,.
\end{align*}
Considering $(\hat{L}^{(m)}_{\mathrm{CP}},0)$ actually makes the computations a little shorter, but we will stick to $(L^{(m)}_{\mathrm{CP}},\overline{\ell}^{(m)}_{\mathrm{CP}})$ as this will facilitate the comparison of our results with the existing literature.

\subsection{Variations}
The variations of the $L^{(m)}_{\mathrm{EH}}$ and $\overline{\ell}^{(m)}_{\mathrm{GHY}}$ terms are standard and they may be found, for instance, in \cite{CPSGR}. The remaining variations are those of the coupling Lagrangians, which can be written as
\begin{align*}
   & \dd L^{(m)}_{\mathrm{CP}}=(\mathfrak{E}^{\mathrm{CP}}_{(m)})^{\alpha \beta} \dd g_{\alpha \beta} + \tensor{(\mathcal{E}^{\mathrm{CP}}_{(m)})}{_\gamma^\alpha^\sigma} \dd \tensor{Q}{^\gamma_\alpha_\sigma} + \mathrm{d} \Theta^{(m)}_{\mathrm{CP}}\,,\\
   & \dd\overline{\ell}^{(m)}_{\mathrm{CP}}-\jmath^*\Theta^{(m)}_{\mathrm{CP}}=0\,,
\end{align*}
where
\begin{align*}
         (\mathfrak{E}^{\mathrm{CP}}_{(m)})^{\alpha \beta}(g,Q)&:= \Big( \tensor{Q}{^\gamma^\alpha_\sigma}\tensor{Q}{^\sigma_\gamma^\beta} - C_{\sigma}\tensor{Q}{^\sigma^\alpha^\beta}  + \frac{1}{2}g^{\alpha \beta}(C_{\sigma}A^{\sigma} - \tensor{Q}{^\gamma^\tau_\sigma}\tensor{Q}{^\sigma_\gamma_\tau})\Big)\vol_{g}\,,\\
     \tensor{(\mathcal{E}^{\mathrm{CP}}_{(m)})}{_\gamma^\alpha^\sigma}(g,Q)&:= \Big( \delta^{\alpha}_{\gamma}A^{\sigma} + g^{\alpha\sigma}C_{\gamma} - \tensor{Q}{^\sigma_\gamma^\alpha} - \tensor{Q}{^\alpha^\sigma_\gamma}\Big) \vol_{g}\,. 
\end{align*}
Notice that the latter is an equation of motion by itself, but the former is not. To obtain one we have to add $\mathfrak{E}^{\mathrm{CP}}_{(m)}$ to the Einstein equations coming from the variation of $L^{(m)}_{\mathrm{EH}}$ (which only depends on $g$).\vspace*{2ex}

In view of the previous result, we can choose the following representatives (defined up to a relative exact form) as the contributions of the $\mathrm{CP}$ terms to the symplectic potentials 
\begin{align*}
     &\Theta^{(m)}_{\mathrm{CP}}(g,Q):= \dd (\imath_{\vec{A}-\Vec{C}}\vol_{\overline{g}})\,,\qquad\qquad\qquad \overline{\theta}^{(m)}_{\mathrm{CP}}(g,Q):=0\,.
\end{align*}

\subsection{Space of solutions}

The algebraic equation of motion $\mathcal{E}^{\mathrm{CP}}_{(m)}(g,Q)=0$ can be solved for $Q$. Its general solution is $Q_0^{\alpha\beta\gamma}=g^{\alpha\gamma}U^\beta$ with arbitrary $U^\beta$ (we sketch the proof of a  similar fact in section \ref{subsect: tetrad SOL}). This solution, moreover, satisfies $\mathfrak{E}^{\mathrm{CP}}_{(m)}(g,Q_0)=0$ for any $g$. Hence, $(g,Q)$ is a solution for Palatini if and only if $Q=Q_0$ and $g$ satisfies the Einstein equations:
  \[\mathrm{Sol}^{(m)}_{\mathrm{PT}}=\{(g_{\alpha\beta},\delta^\alpha_\gamma U_{\!\beta})\ /\ g\in\mathrm{Sol}^{(m)}_{\mathrm{GR}},\ U_{\!\beta}\text{ arbitrary}\}\,.\]
  This, in turn, proves that the metric sector of Palatini  is equivalent to metric-GR. Notice that the boundary only plays a role in the metric sector  $\mathrm{Sol}^{(m)}_{\mathrm{GR}}$ of the solution space (which is discussed in detail in \cite{CPSGR}, where both Dirichlet and Neumann boundary conditions are considered). We  have now the following (``on shell'') identities over the space of solutions 
   \begin{equation}\label{eq: tilde=LC+}
   \begin{array}{ll}
       \mathrm{\widetilde{R}}\tensor{\mathrm{iem}}{^\alpha_\beta_\mu_\nu}=\mathrm{\LC{R}}\tensor{\mathrm{iem}}{^\alpha_\beta_\mu_\nu}+g^\alpha_\beta(\d U)_{\mu\nu}\,,\\[0.3ex]
        \mathrm{\widetilde{R}ic}_{\beta\nu}=\mathrm{\LC{R}ic}_{\beta\nu}+(\d U)_{\beta\nu}\,,&  \widetilde{R}=\LC{R}\,,\\
        \widetilde{K}_{\overline{\alpha}\overline{\beta}}=\LC{K}_{\overline{\alpha}\overline{\beta}}\,,&  \widetilde{K}=\LC{K}\,,\\ \widetilde{M}_{\alpha\beta\gamma}=-2 g_{\beta\gamma}U_\alpha\,,  &  \tensor{\widetilde{T}}{^\gamma_\alpha_\beta}=\delta^\gamma_\beta U_\alpha-\delta^\gamma_\alpha U_\beta\,.
    \end{array}
    \end{equation}
The last two equations imply that, on solutions, $\widetilde{M}=0$ if and only if $\widetilde{T}=0$. Thus, we have either the Levi-Civita connection or one with nonmetricity and torsion.

\subsection{Presymplectic form}
     The metric-Palatini presymplectic form on the space of solutions $\OOmega^{\mathrm{PT}}_{(m)}$ has the same functional form as the metric-GR presymplectic form $\OOmega^{\mathrm{GR}}_{(m)}$ since $\dd\Theta^{(m)}_{\mathrm{CP}}=0$. 
     However, the solution spaces are different. In order to compare them, we need the projection $\pi_{(m)}(g,Q)=g$. In this way we get
     \begin{equation}\label{eq: O_PT(m)=pi Omega_GR(m)}
         \OOmega^{\mathrm{PT}}_{(m)}=\pi_{(m)}^*\OOmega^{\mathrm{GR}}_{(m)}
     \end{equation} 
Clearly, all vectors of the form $(0,\mathbb{V})\in T\mathrm{Sol}^{(m)}_{\mathrm{PT}}$, i.e. in the directions of the connection sector, define gauge directions since they belong to the kernel of $(\pi_{(m)})_*$ and hence annihilate $\OOmega^{\mathrm{PT}}_{(m)}$.

\section{\label{sec:TETRADPALATINI}TETRAD PALATINI}

Given a tetrad $e_\alpha^I$, we have its $\eta_{IJ}$-dual $E^\alpha_I$ and the space-time metric $g=\eta_{IJ}e^Ie^J=:\Phi_{\mathrm{GR}}(e)$ as explained in more detail in \cite{CPSGR}. We can consider the Levi-Civita connection $\LC{\nabla}$ of $g$ which, in turn, allows us to build the $1$-form connection
\begin{align}\label{eq: LC omega}
\tensor{\LC{\omega}}{_\mu^K_I}:= e^{K}_{\alpha}\LC{\nabla}_{\!\mu}E^{\alpha}_{I}\,.
\end{align}
If we now consider a generic $1$-form connection $\tensor{\widetilde{\omega}}{_\mu^I^J}$ (not necessarily antisymmetric in the internal indices), we can define the covariant derivative
\begin{align*}
    \widetilde{\nabla}_{\!\mu}Y^\beta=(\mathrm{d} Y^J)_\mu E^\beta_J+ E^\beta_K\tensor{\widetilde{\omega}}{_\mu^K_J}Y^J\,, \qquad Y^J:=e^J_\gamma Y^\gamma\,,
\end{align*}
and extend it in the usual way to objects with more space-time and internal indices. In particular, using this definition with $E^\beta_I$ leads to
\begin{equation}
\tensor{\widetilde{\omega}}{_\mu^K_I} = e^{K}_{\alpha}\widetilde{\nabla}_{\!\mu}E^{\alpha}_{I}\,,
\end{equation}
which is analogous to \eqref{eq: LC omega}.  We also have the  $\widetilde{\omega}$-curvature and the $\widetilde{\omega}$-covariant derivative over forms given by
\begin{align*}
&\widetilde{F}_{IJ} = \d \widetilde{\omega}_{IJ} + \widetilde{\omega}_{IK}\wedge \tensor{\widetilde{\omega}}{^K_J}\,,
&\widetilde{\mathcal{D}}\alpha_I=\mathrm{d}\alpha_I+\tensor{\widetilde{\omega}}{_I^J}\wedge\alpha_J\,,
\end{align*}
where the latter is extended, as usual, to objects with more antisymmetric internal indices.
 
Given two 1-form connections $\omega$ and $\widetilde{\omega}$, we have two covariant derivatives $\nabla$ and $\widetilde{\nabla}$ which are related by a $(2,1)$-tensor $Q$. In particular, if we take $\nabla=\LC{\nabla}$, then
\begin{equation*}
    \tensor{Q}{^\beta_\mu_\alpha}=E^\beta_Ke^J_\alpha\big(\tensor{\widetilde{\omega}}{_\mu^K_J}-\tensor{\LC{\omega}}{_\mu^K_J}\big)=:\tensor{\varphi(e,\widetilde{\omega})}{^\beta_\mu_\alpha}\,.
\end{equation*}
This allows us to define the map
\[\Phi_{\mathrm{PT}}(e,\widetilde{\omega}):=\big(\Phi_{\mathrm{GR}}(e),\varphi(e,\widetilde{\omega})\big)\]
which is surjective but not injective. In fact $\Phi_{\mathrm{PT}}(e,\widetilde{\omega})=\Phi_{\mathrm{PT}}(e',\widetilde{\omega}')$ if and only if there exists some $\Psi\in SO(1,3)$ such that $e'_I=\tensor{\Psi}{_I^J}e_J$ and $\widetilde{\omega}_I ^{'\,J}=\tensor{\Psi}{_I^K}\d\tensor{\Psi}{^J_K}+\tensor{\Psi}{_I^K}\tensor{\Psi}{^J_L}\widetilde{\omega}\tensor{}{_K^L}$.
\subsection{The action}
In order to introduce the tetrad formalism, we perform the ``change of variables'' given by $\Phi_{\mathrm{PT}}$
\begin{align*}
    &L^{(t)}_{\mathrm{PT}}(e, \widetilde{\omega}) :=L^{(m)}_{\mathrm{PT}}\circ\Phi_{\mathrm{PT}}(e, \widetilde{\omega})\,,&\overline{\ell}^{(t)}_{\mathrm{PT}}(e, \widetilde{\omega}) :=\overline{\ell}^{(m)}_{\mathrm{PT}}\circ\Phi_{\mathrm{PT}}(e, \widetilde{\omega})\,.
\end{align*}
We could write again these Lagrangians as the GR part plus a coupling term, but in this case it is more convenient to split the $1$-form connection in its antisymmetric and symmetric parts (in its internal indices) 
\begin{equation*}
    \widetilde{\omega}_{IJ} = \widehat{\omega}_{IJ} + S_{IJ}\,,
\end{equation*}
and consider the equivalent variables $(e,\widehat{\omega},S)$ instead. Following the ideas of \cite{CPSGR}, one obtains the following expressions for these Lagrangians
\begin{align*}
    L^{(t)}_{\mathrm{PT}}(e, \widehat{\omega}, S) &= \frac{1}{2}\epsilon_{IJKL}\Big( \widehat{F}^{IJ}- \frac{\Lambda}{6}e^{I} \wedge e^{J}+\tensor{S}{^I_M}\wedge S^{MJ} \Big) \wedge e^K\wedge e^L\,,\\
    \overline{\ell}^{(t)}_{\mathrm{PT}}(e, \widehat{\omega})&= -\frac{1}{2}\epsilon_{IJKL}\left(2 N^{I}\d N^{J} - \overline{\widehat{\omega}}{}^{IJ}\right)\wedge \overline{e}^K\wedge \overline{e}^{L}\,,
\end{align*}
where $\overline{e}^I:=\jmath^*e^I$, $\overline{\widehat{\omega}}{}^{IJ}:=\jmath^*\widehat{\omega}^{IJ}$ and $N^I=\nu^\alpha e^I_\alpha$. Notice that  $\overline{\ell}^{(t)}_{\mathrm{PT}}$ does not depend on $S$.

\subsection{Variations}
Computing the variations, one easily obtains
\begin{align*}
    & \dd L^{(t)}_{\mathrm{PT}} = \mathfrak{E}^{(t)}_{L}\wedge \dd e^{L} + \mathcal{E}^{(t)}_{KL}\wedge \dd \widehat{\omega}^{KL}+ \mathscr{E}^{(t)}_{JM}\wedge \dd S^{JM}  + \d \Theta^{(t)}_{\mathrm{PT}}\,, \\
   & \dd \overline{\ell}^{(t)}_{\mathrm{PT}} - \jmath^{*}\Theta^{(t)}_{\mathrm{PT}} = \overline{b}^{(t)}_{I}\wedge \dd \overline{e}^{I} - \d \overline{\theta}^{(t)}_{\mathrm{PT}}\,,
\end{align*}
where the Euler-Lagrange equations are
\begin{align*}
    \mathfrak{E}^{(t)}_{L} &:= \epsilon_{IJKL}\Big( \widehat{F}^{IJ} +\tensor{S}{^I_M}\wedge S^{MJ}- \frac{\Lambda}{3} e^{I} \wedge e^{J}\Big) \wedge e^{K}\,,\\
    \mathcal{E}^{(t)}_{KL} &:= \frac{1}{2}\widehat{D}(\epsilon_{IJKL}e^{I}\wedge e^{J})\,,\\
    \mathscr{E}^{(t)}_{JM}&:= \frac{1}{2}\big( \epsilon_{IKLJ}\delta_M^R + \epsilon_{IKLM}\delta_J^R\big)\tensor{S}{_R ^I}\wedge e^{K} \wedge e^{L}\,,\\
     \overline{b}^{(t)}_{I} &:= \epsilon_{IJKL}(2N^{K}\d N^{L} - \overline{\widehat{\omega}}{}^{KL})\wedge \overline{e}^{J} +2\epsilon_{MJKL}N^L(\iota_{\overline{E}^J}\d\overline{e}^K)\wedge\overline{e}^MN_I \,,
\end{align*}
and we take the symplectic potentials
\begin{align}
    &\Theta^{(t)}_{\mathrm{PT}} := \frac{1}{2}\epsilon_{IJKL}e^{I}\wedge e^{J} \wedge \dd \widehat{\omega}^{KL}\,,\label{eq: Theta PT}
    &\overline{\theta}^{(t)}_{\mathrm{PT}} := \epsilon_{IJKL}\overline{e}^{I}\wedge \overline{e}^{J}  \wedge N^{K} \dd N^{L}\,.
\end{align}
Notice that the boundary term $\overline{b}_I$ plays no role in the Dirichlet case as in the variational principle one has $\dd \overline{e}^I=0$, see \cite{CPS} for a careful discussion.

\subsection{Space of solutions}\label{subsect: tetrad SOL}
As in the metric case, we can exactly solve some of the equations. We begin by expanding  $\tensor{S}{_I_J} = \tensor{S}{_I_J_K} e^{K}$ and using the unique decomposition $\tensor{S}{_I_J_K} = \slashed{S}_{IJK} + \eta_{IJ}U_{K}$ with $\slashed{S}\tensor{}{^I_I_K}=0$. Plugging this into $\mathscr{E}^{(t)}=0$ we get
\begin{align*}
        &\tensor{\slashed{S}}{_R_L_I} + \tensor{\slashed{S}}{_I_L_R} - \slashed{S}\tensor{}{_R_J^J}\eta_{LI} - \slashed{S}\tensor{}{_I_J^J}\eta_{LR} = 0\,,
\end{align*}
and taking its trace shows that $\slashed{S}\tensor{}{_I_J^J}=0$ which, in turn, implies  $\slashed{S}_{IJK}=0$. The general solution to $\mathscr{E}^{(t)}=0$ is then $S_{IJK} = \eta_{IJ}U_{K}$, with arbitrary $U_{\!K}$. Similarly, from  $\mathcal{E}^{(t)}=0$ we obtain the solution $\widehat{\omega}_{IJ}=\LC{\omega}_{IJ}$. Thus
  \[\mathrm{Sol}^{(t)}_{\mathrm{PT}}=\{(e^I_\alpha,\LC{\omega}_\mu^{IJ}+\eta^{IJ}U_{\!\mu})\ /\ e^I_\alpha\in\mathrm{Sol}^{(t)}_{\mathrm{GR}},\ U_{\!\mu}\text{ arbitrary}\}\]
  We see, again, that the tetrad sector of Palatini is equivalent to tetrad-GR and that the boundary plays no role in the connection part. As in the metric case, the Dirichlet or Neumann boundary conditions for the tetrads are incorporated in $\mathrm{Sol}^{(t)}_{\mathrm{GR}}$, which is studied in detail in \cite{CPSGR}.

\subsection{Presymplectic form}
It is easy to see that, over the space of solutions, both symplectic potentials on \eqref{eq: Theta PT}  are given by the same expressions as the symplectic potentials obtained in  \cite{CPSGR} for tetrad GR (remember, though, that they live in different spaces). In fact, using the projection $\pi_{(t)}(e,\widetilde{\omega})=e$ we have
\begin{equation}\label{eq: O_PT(t)=pi Omega_GR(t)}
         \OOmega^{\mathrm{PT}}_{(t)}=\pi_{(t)}^*\OOmega^{\mathrm{GR}}_{(t)}
     \end{equation}
in analogy with equation \eqref{eq: O_PT(m)=pi Omega_GR(m)}. We conclude as well that the vectors of the form $(0,\mathbb{W})\in T\mathrm{Sol}^{(t)}_{\mathrm{PT}}$ correspond to degenerate directions of $\OOmega^{\mathrm{PT}}_{(t)}$.

\section{\label{sec:CONCLUSIONS}CONCLUSIONS AND COMMENTS}
In this letter, we have studied Palatini gravity, both in the metric and tetrad formulations, in a manifold with boundary, and including nonmetricity and torsion. As can be seen in \eqref{eq: O_PT(m)=pi Omega_GR(m)} and \eqref{eq: O_PT(t)=pi Omega_GR(t)}, the presymplectic structures defined on the space of solutions of metric and tetrad Palatini are related to those of metric and tetrad GR by the projection on the first factor. Also, the solution spaces of metric and tetrad Palatini are related to each other by $\Phi_{\mathrm{PT}}$.  Finally, since we have from \cite{CPSGR} that
\begin{equation}\label{eq: O_GR(t)=Phi Omega_GR(m)}
 \OOmega^{\mathrm{GR}}_{(t)}=\Phi^*_{\mathrm{GR}}\OOmega^{\mathrm{GR}}_{(m)}
\end{equation}
it is immediate to obtain, using  $\pi_{(m)}\circ\Phi_{\mathrm{PT}}=\Phi_{\mathrm{GR}}\circ\pi_{(t)}$, the following important relation
\begin{equation}\label{eq: O_PT(t)=Phi Omega_PT(m)}
 \OOmega^{\mathrm{PT}}_{(t)}=\Phi^*_{\mathrm{PT}}\OOmega^{\mathrm{PT}}_{(m)}
\end{equation}
This is one of the main results of the paper: the equivalence, up to the internal gauge transformations given by the kernel of $(\Phi_{\mathrm{PT}})_*$, of metric and tetrad Palatini. Furthermore, equations \eqref{eq: O_PT(m)=pi Omega_GR(m)}, \eqref{eq: O_PT(t)=pi Omega_GR(t)}, \eqref{eq: O_GR(t)=Phi Omega_GR(m)}, and \eqref{eq: O_PT(t)=Phi Omega_PT(m)} show the equivalence (up to the gauge transformations given by the kernel of the corresponding push-forwards) of all these four formulations of gravity.\mbox{}\vspace*{-2ex}

\begin{equation}
\begin{tikzpicture}[baseline=(current  bounding  box.center)]
  \matrix (m) [matrix of math nodes,row sep=1.5em,column sep=4em,minimum width=2em]
  {
     \big(\OOmega_{(t)}^{\mathrm{PT}},\mathrm{Sol}_{(t)}^{\mathrm{PT}}\big) & \big(\OOmega_{(m)}^{\mathrm{PT}},\mathrm{Sol}_{(m)}^{\mathrm{PT}}\big) \\
     \big(\OOmega_{(t)}^{\mathrm{GR}},\mathrm{Sol}_{(t)}^{\mathrm{GR}}\big) & \big(\OOmega_{(m)}^{\mathrm{GR}},\mathrm{Sol}_{(m)}^{\mathrm{GR}}\big) \\};
  \path[-stealth]
    (m-1-1) edge node [left] {$\pi_{(t)}$} (m-2-1)
    (m-1-1) edge node [above] {$\Phi_{\mathrm{PT}}$} (m-1-2)
    (m-2-1) edge node [below] {$\Phi_{\mathrm{GR}}$} (m-2-2)
    (m-1-2) edge node [right] {$\pi_{(m)}$} (m-2-2);
\end{tikzpicture}
\end{equation}
This has been proven for both Dirichlet and homogeneous Neumann boundary conditions, although the same result can be obtained in other instances as long as all four formulations are related in the way shown in the diagram (hence with the appropriate boundary Lagrangians, \cite{CPSGR}). Notice that this may also be seen by realizing that the projection of the spaces of solutions over the metric/tetrad sector is well defined (in the sense that it is not necessary to know the second factor to get the first). On the other hand, the main result regarding the connection sector is that boundaries have no influence whatsoever (once the correct boundary Lagrangian is considered). We have also found the well-known projective symmetry in both metric and tetrad formulations, which can be expressed, respectively, in the form 
\begin{equation}
    \tensor{Q}{^\alpha_\beta_\gamma}=\delta^\alpha_\gamma U_{\!\beta}\,,\qquad\qquad S^{IJ}_\beta=\eta^{IJ}U_{\!\beta}
\end{equation}
for arbitrary $U_{\!\beta}$. This projective symmetry plays an important role to define observables as they must be $U$-independent. For instance, the $g$-Kretschmann curvature $\mathrm{\LC{K}res}:=\mathrm{\LC{R}iem}_{\alpha\beta\gamma\delta}\mathrm{\LC{R}iem}{}^{\alpha\beta\gamma\delta}$ leads to an observable for every $(g,Q)$ as, in fact, it is independent of $Q$. However, the $(g,Q)$-Kretschmann $\widetilde{\mathrm{K}}\mathrm{res}:=\mathrm{\widetilde{R}iem}_{\alpha\beta\gamma\delta}\mathrm{\widetilde{R}iem}{}^{\alpha\beta\gamma\delta}$ cannot provide an observable. Indeed, using \eqref{eq: tilde=LC+}, it is easy to show that on the space of solutions we have 
\[\widetilde{\mathrm{K}}\mathrm{res}=\mathrm{\LC{K}res}+(\d U)_{\alpha\beta}(\d U)^{\alpha\beta}\]
which depends on $U$. This is important if one plans to use these geometric objects as indicators of the presence of singularities \cite{Bejarano2020}.

Finally, it is relevant to mention that although we have proven the equivalence of the four theories in vacuum, this equivalence may be broken by matter fields. For instance, pregeodesics computed by using the connection are insensitive to the $U$-arbitrariness  \cite{Bernal2017}. However, if matter is introduced, this may no longer be the case. Since we consider Palatini theories where both nonmetricity and torsion are allowed to be non-zero, matter fields can be coupled in ways that are not possible in standard GR.  This may produce interesting changes in the dynamics of the gravity-matter system.

\begin{acknowledgments}
The authors are grateful to  G. Olmo for helpful discussions.
This work has been supported by the Spanish Ministerio de Ciencia Innovaci\'on y Uni\-ver\-si\-da\-des-Agencia Estatal de Investigaci\'on FIS2017-84440-C2-2-P grant. Juan Margalef-Bentabol is supported by the Eberly Research Funds of Penn State, by the NSF grant PHY-1806356 and by the Urania Stott fund of Pittsburgh foundation UN2017-92945. E.J.S. Villase\~nor is supported by the Madrid Government (Comunidad de Madrid-Spain) under the Multiannual Agreement with UC3M in the line of Excellence of University Professors (EPUC3M23), and in the context of the V PRICIT (Regional Programme of Research and Technological Innovation).
\end{acknowledgments}

\bibliographystyle{plainnat}
\bibliography{PRL}
\end{document}